\documentstyle[11pt,proceedings,twoside,epsf]{article}
\begin{document}

\title{The Tail of the HI Mass Function}
\author{Stephen E. Schneider \& Jessica L. Rosenberg}
\affil{University of Massachusetts, Department of Astronomy, Lederle
Graduate Tower, Amherst, MA 01003}

\begin{abstract}
The contribution of extragalactic objects with HI masses below $10^8 M_\odot$
to the HI mass function remains uncertain. Several aspects of the detection of
low-mass sources in HI surveys are not always considered, and as a result
different analysis techniques yield widely different estimates for their number
density. It is suggested at one extreme that the number density of galaxies
follows a shallow Schechter power-law slope, and at the other extreme that it
follows a steep faint-end rise like that found for field optical sources. Here
we examine a variety of selection effects, issues of completeness, and
consequences of LSS. We derive results for the large Arecibo
Dual Beam Survey which indicate that the field mass function does rise steeply,
while within the Virgo Cluster environs, the slope appears to be much
shallower. Dependence on the local density of galaxies may partially explain
differences between surveys.
\end{abstract}

\section{Introduction}

The shape of the HI mass function, particularly the behavior below $10^8
M_\odot$, has been the subject of considerable controversy. As paramaterized by
a Schechter function slope $\alpha$, several studies seem to indicate a fairly
shallow slope of $\alpha\approx-1.2$, including the HI masses of: optically-selected samples
(e.g., Briggs \& Rao 1993; Huchtmeier, this volume); galaxies in high-density
regions like the Canes Venatici group (Kraan-Korteweg et al.~1999), Centaurus A
(Banks et al. 1999), and the Ursa Major cluster (Verheijen et al.~this volume);
even some ``blind'' (optically unbiased) surveys.

On the other hand, several studies suggest the slope may be considerably
steeper in field samples. Initial reports at this meeting of results from the
Parkes survey suggest a slope of $\alpha=-1.5$ (Webster et al.~this volume).
Interestingly, Kraan-Korteweg et al. (1999) found $\alpha=-1.4$ when they did
not restrict their sample to the Canes Venatici group. Our analysis of two
earlier Arecibo surveys (Schneider et al. 1998; SSR hereafter) also suggested a
steeper slope or a faint-end rise similar to the rise found for an optical
sample of field objects found by Loveday (1997) and Driver \& Phillipps (1996).

There are several possible causes for different conclusions to be drawn about
the population of low-mass HI sources. Some of these have to do with
astrophysical characteristics of the sample, like the effects of LSS
and distance uncertainties. However, a significant cause of the
differences is analysis problems: (1) a poor understanding of survey
sensitivity; (2) small number statistics; and (3) selection effects.

To address these problems, we have completed the largest blind HI survey to
date at the Arecibo Observatory. The Arecibo Dual-Beam Survey (ADBS; Rosenberg
\& Schneider 2000 and this volume) covers 430 sq. deg. in drift scans at 30
separate declinations between  $9^\circ < \delta < 28^\circ$. We were able to
achieve an rms sensitivity of 3-4 mJy per 3.3 arcmin beam along the drift
scans. We also repeated most of the scans to provide internal confirmation of
source detections. The advantages of using Arecibo for this type of survey are
discussed by Rosenberg \& Schneider (2000).

The most important new feature of the ADBS is the insertion of hundreds of
``synthetic'' HI signals prior to applying the data reduction procedures. These
sources allow us to test  our actual sensitivity and completeness directly. We
discuss these issues in detail in \S 2. The resulting mass function for 
the field population of galaxies in the ADBS is derived in \S 3.

Large-scale structure (LSS) may also have an effect on the shape of the mass
function. Density differences also affect the derivation of the mass function,
as do large scale flows and velocity dispersion relative to the Hubble flow. We
examine these potential effects in \S 4. Our analysis of the ADBS suggests that
field samples of HI-selected sources {\it do} show a steep slope, while
optically-selected samples and cluster samples show a shallower slope. We
conclude with a discussion of these results and future prospects in \S 5.

\section{Previous Arecibo Surveys and the Importance of Completeness}

Determining the sensitivity limit of an HI survey is, in many ways, more
critical than the size of the survey. If the wrong limit is used, it will
systematically bias the mass function. The term ``sensitivity limit'' is itself
a misnomer: the completeness changes as a function of both signal
strength and line width, and it is {\it not} a sharp cut at some ``$n$-$\sigma$''
threshold, as is assumed in many studies.

For two earlier Arecibo HI surveys (Zwaan et al. 1997; Spitzak \& Schneider
1998) a $V/V_{max}$ test indicates that the samples are not complete to
``5--$\sigma$'' as is traditionally assumed (SSR). In particular, the Zwaan et
al.~sample did not satisfy the $V/V_{max}$ test unless a significantly higher
cutoff level was assumed. This higher effective noise level means that Zwaan et
al.~were sensitive to low mass sources within a smaller volume than originally
claimed, and therefore the density of these sources is higher than they
determined. By contrast, the volume within which high mass sources were
detectable is bandpass limited, so the net effect of underestimating the
limiting sensitivity is to suppress the steepness of the mass function.

In addition, the incompleteness of the two surveys did not follow the usual
assumption that noise scales as the line width as $w^{0.5}$, as is expected
from simple statistical arguments (e.g., Schneider 1996). Both Arecibo samples
showed a tendency to be less complete for wider-line profiles with an effective
cutoff $\propto w^{0.75}$. This width dependence implies a lower completeness
for wide-lined galaxies. A likely source of this problem is the baseline
subtraction. Both automated methods and visual examination generally identify
which channels to mask from the baseline fit by their deviation from
neighboring channels. When the signal is spread over more channels, the
difference from neighboring channels is reduced, and the channels may not be
masked. Since wide-line signals are usually from HI-massive galaxies, this will
suppress the counts at the high mass end, although the effect is usually minor
since high-mass sources tend to be bandpass limited.
Applying a $w^{0.75}$ cutoff, both samples passed the $V/V_{max}$ test and were
consistent with each other.

For these earlier Arecibo samples, we assumed a sharp cutoff at the ratio of
signal-to-noise ($S/N$) that passed the $V/V_{max}$ test. To determine the
completeness as a function of $S/N$ in the ADBS, we generated a set
of synthetic signals with random positions, line-widths, and line-strengths,
and inserted them into the observing database. The sources had profile shapes
that were modeled to look very similar to typical galaxy HI line profiles. We
included sources ranging from several times weaker to several times stronger
than a first-guess at our typical detection limit. These fake signals were
added to the ADBS spectra before any of the ``baselining'' steps were
performed. We were then able to test how efficiently our recovery process
worked so that we could characterize the probable fraction of sources we were
missing for any line width and $S/N$ value.

\begin{figure}
\plottwo{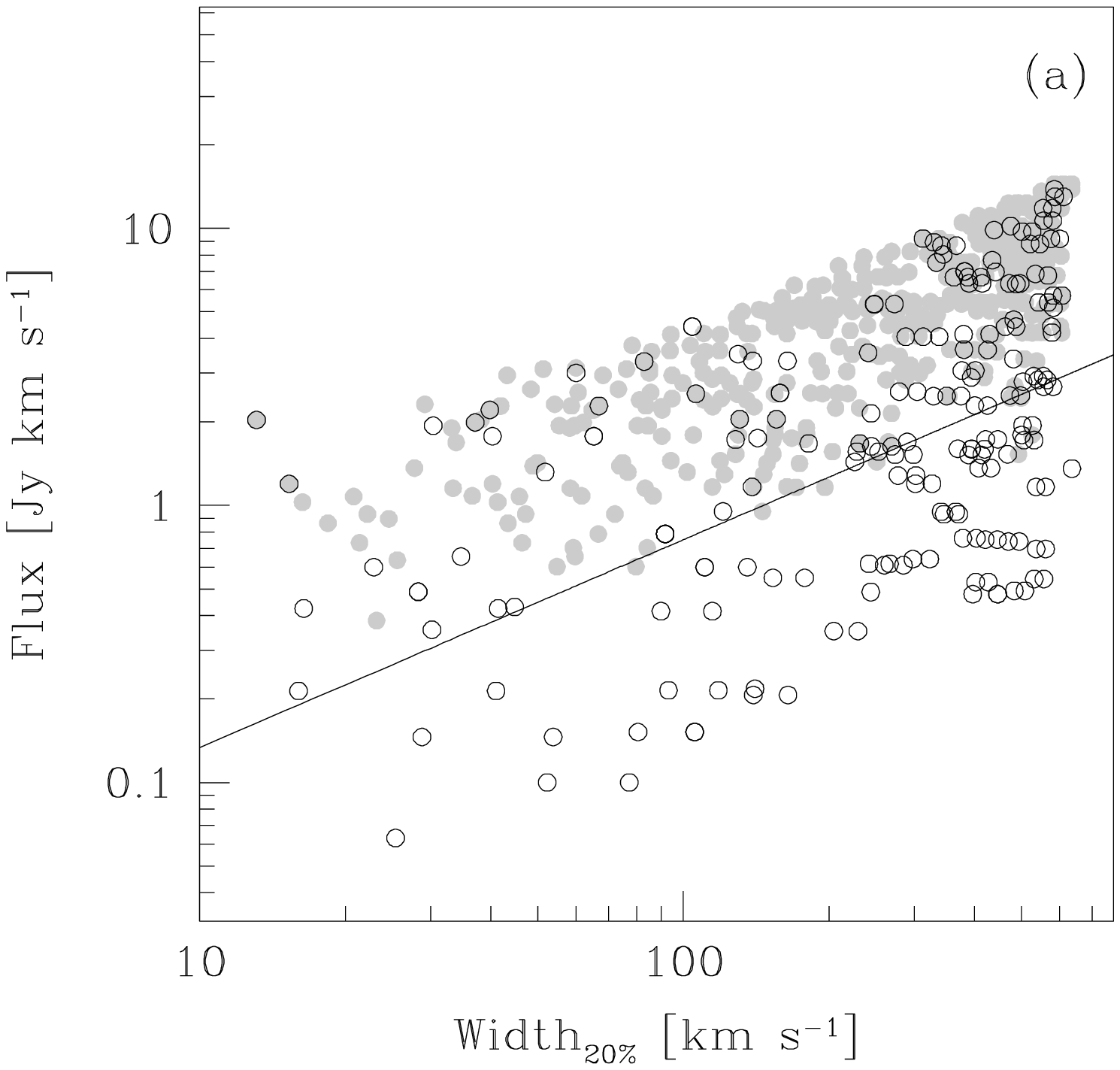}{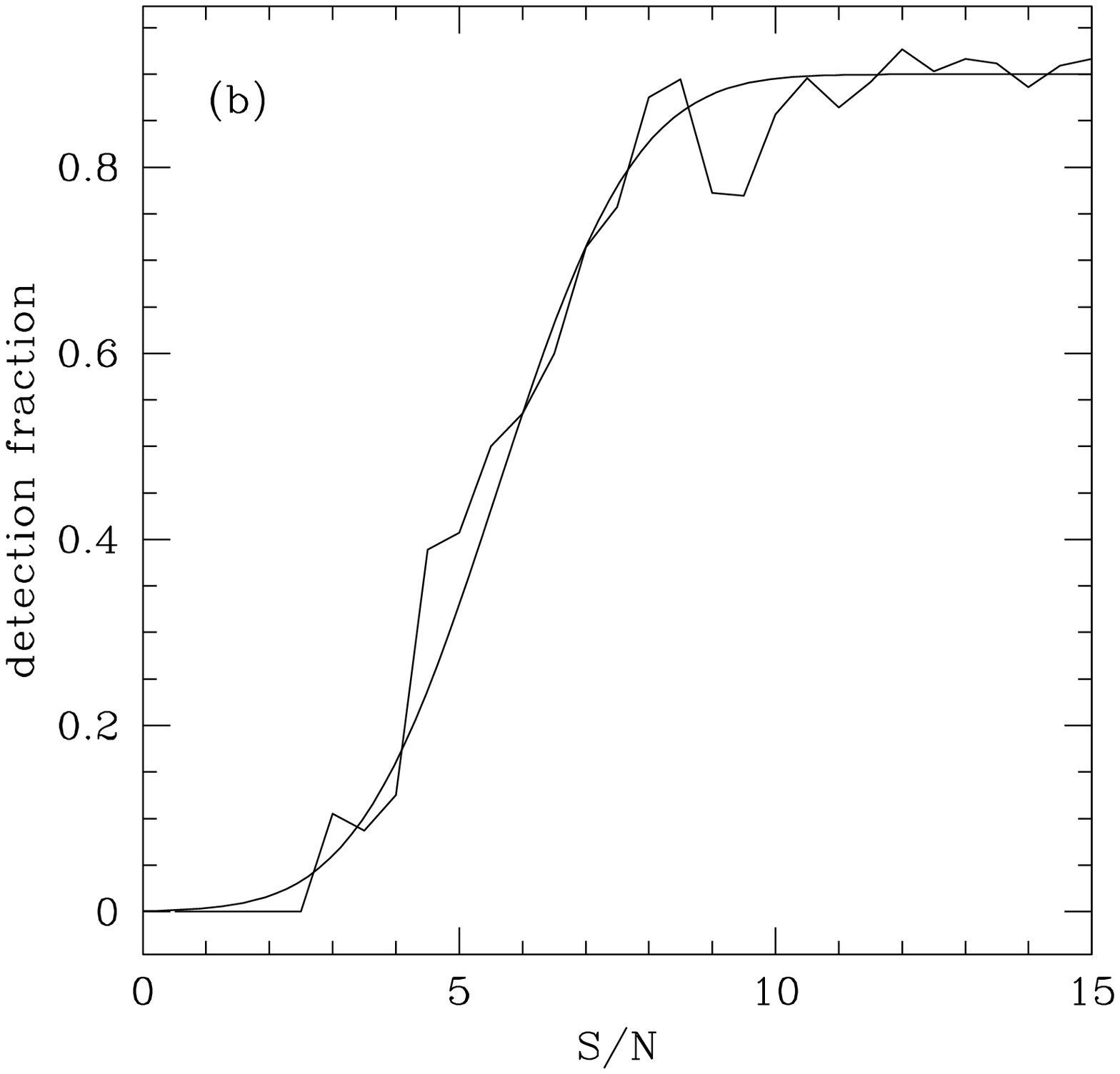}
\caption{(a) A plot showing the fluxes and line widths of synthetic sources.
Solid gray circles indicate sources that were recovered---note that these
points run together, so that the large number of them is not obvious. The
open circles indicate sources that were missed. The diagonal line shows a
width$^{0.75}$ dependence. (b) The completeness fraction as a function
of the effective signal-to-noise ratio. The smooth curve shows an error function
fit to the results for the synthetic sources.} 
\end{figure}

In Fig.~1a, we show the fluxes and line widths of the synthetic sources,
indicating the undetected sources by open circles. The plot is a little
deceptive at the wide-line end of the plot in that there are many more detected
sources (shown as gray dots) than are visible in the plot because the dots run
together. The diagonal line in the figure has a slope $\propto w^{0.75}$, which
is a good description of the boundary below which very few sources are
detected.

In Fig.~1b, we show the fraction of recovered synthetic sources as a function
of $S/N$, where we have folded together sources with
different line widths by dividing first by a value $\propto w^{0.75}$.  The
value of S/N is scaled so that it gives the normal value of $S/N$ for
a source 300 km s$^{-1}$ wide (see SSR). The shape of the resulting
completeness curve is quite similar to an error function. This is what one
would expect for Gaussian noise added to an underlying signal---some sources
are pushed below the threshold for detection while others are pushed above it.

Note that about 10\% of high $S/N$ synthetic sources were not
recovered. These sources generally fell on top of local Galactic HI emission, 
radio interference, or bandpass edges, making detection difficult. In the final
calibration of the space density of these sources, this is an important
adjustment, but it does not affect the slope of the mass function.

\section{The $V_{tot}$ Method and Resulting Mass Functions}

Using the completeness function derived above, we can accurately determine the
effective search volume of our survey for each detected source.  We do this by
determining the distance to which a source could have been detected as a
function of (1) beam offset, (2) rms noise at each observed position, and (3)
frequency dependence of telescope gain. We then integrate over all of these
possibilities, weighting the volume by the completeness function, to derive the
total effective volume within which each source {\it might} have been detected,
$V_{tot}$. (This is sometimes called $V_{max}$ but this is confusing because it
is different than the value used in the $V/V_{max}$ test.) We have tested the
$V_{tot}$ method quite extensively with simulations, and find that it does an
excellent job of recovering the input mass function.

Weighting the volume elements by the completeness function solves the problem
of dealing with sources near the completeness roll off. For example, for each
source that is detected at the 50\% completeness point, a second one is missed.
Therefore the space density of these sources is twice as high as the detected
source implies, or, effectively, half as much volume was searched at the
distance where the source becomes this weak. 

\begin{figure}
\plottwo{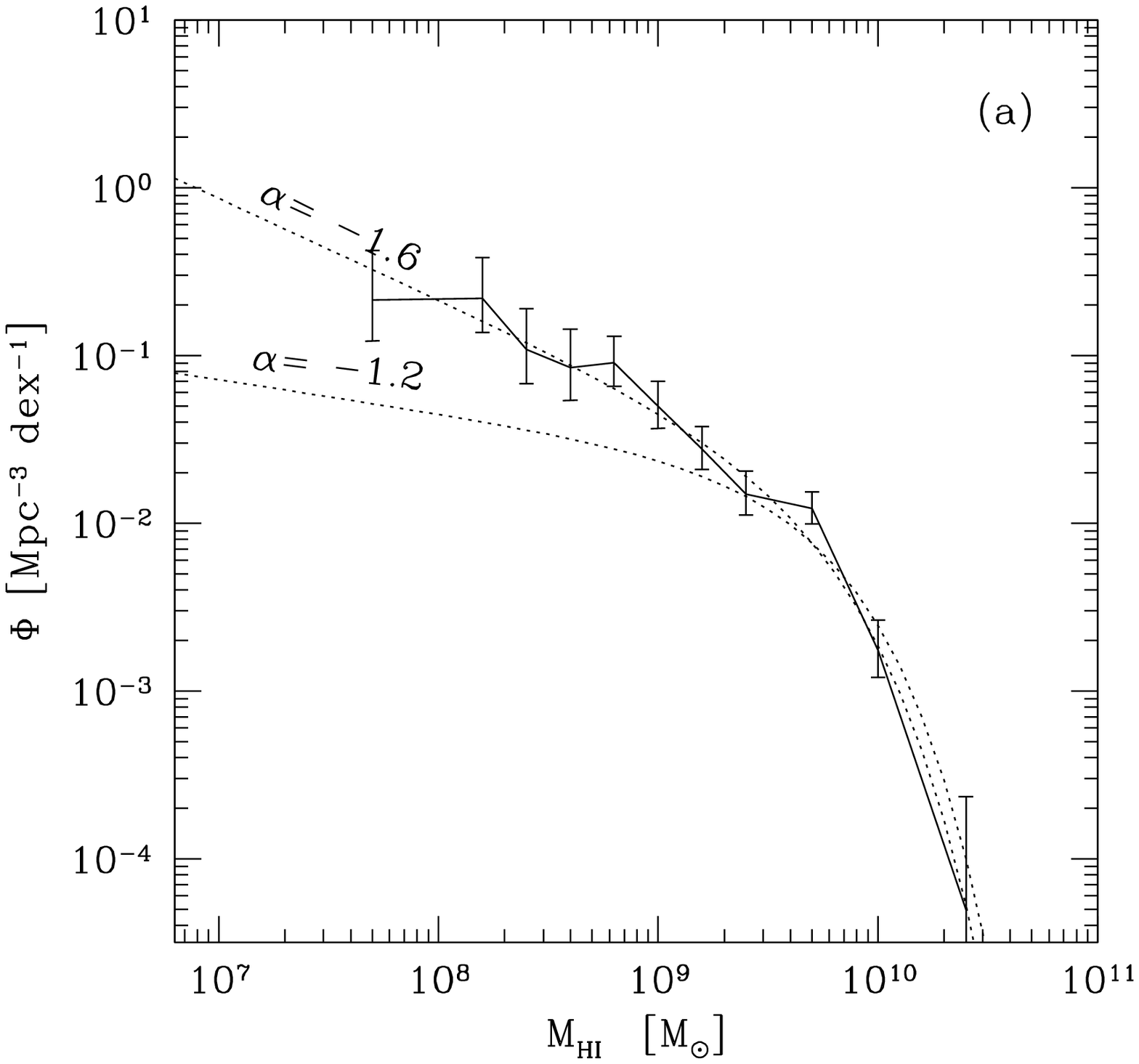}{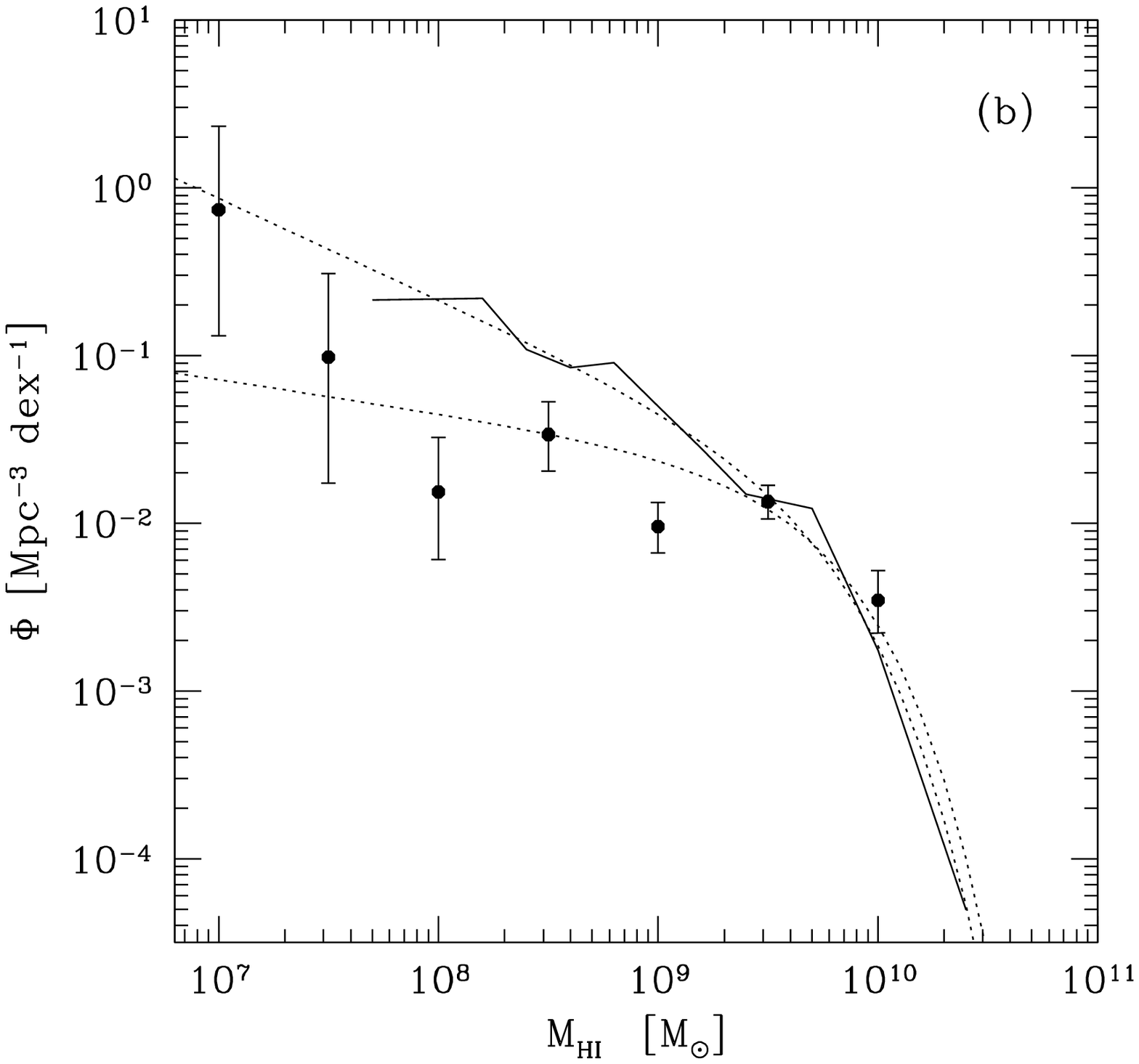}
\caption{(a) The field HI mass function based on the ADBS. (b) Comparison of
the ADBS field mass function with the measurements from two earlier Arecibo HI
surveys (SSR). In both figures the error bars show 95\% confidence intervals,
and the two dotted curves show Schechter functions with slopes of $\alpha=-1.2$
and $-1.6$.} 
\end{figure}

The resulting mass function of the ADBS field galaxies is shown in Fig.~2a.
This function is based only on galaxies farther from the center of the Virgo
Cluster than 27$^\circ$ as we discuss in the next section. The field mass
function displays a fairly good match to a simple Schechter function with a
slope of $\alpha=-1.6$. A Schechter function with $\alpha=-1.2$ sits well below
our curve. Note that the points are unequally spaced because they were grouped
in intervals to maintain at least 6 sources per point, except the highest mass
bin, which includes only 1 source. The error bars show the 95\% confidence
interval based on Poisson statistics for small numbers of sources (Gehrels
1986).

To match the shallower sloped Schechter function, the effective search volume
would have to be $>2\times$ larger for sources with masses $<10^9 M_\odot$,
rising to $\sim5\times$ larger for sources with masses $<10^8 M_\odot$. This
would require that the survey sensitivity is $\sim1.7\times$ smaller than
actually measured (once bandpass limits are properly accounted for). Such a
large effective search volume is unsupported by our completeness measurements
and would make our sample fail the $V/V_{max}$ test. 

We also tested our completeness against sources from the Zwaan et al.~(1997)
23$^\circ$ strip, which one of our driftscans nearly overlapped. Their survey
integrated about 20--30 times longer in the area covered, so that they provide
a good check of our completeness, albeit within a relatively small area. We
detected six of their sources with S/N as low as 5.1, and detected none of the
19 others with S/N as high as 3.1. Unfortunately, there are very few sources in
the range where we might test the shape of the completeness roll off.

In Fig.~2b, we show the measurements from our analysis of the two earlier
Arecibo blind surveys. These earlier data suggested a fairly shallow slope that
begins rising below about $10^8 M_\odot$ of HI. The ADBS mass function is in
reasonable agreement at the high and low mass ends, but the middle range is
significantly lower in the previous samples. We suspect this difference may
reflect a dependence of the mass function associated with LSS, as we discuss
next.

\section{The Effects of Large Scale Structure}

As we showed in our analysis of the earlier Arecibo samples (Schneider
et al.~1998), corrections for density variations along the line of
sight due to LSS are generally small. We based this
on the density variations of optically-selected sources, which are
thought to be more concentrated to high-density regions than HI-selected
sources, so that these were probably over-corrections.

However, such density corrections assume that the {\it
shape} of the mass function is independ of density. Otherwise, the density
effects will be reflected in different ways depending on the survey sensitivity
and the distances of any significant density structures along the line of
sight.

Based on the distribution of optically selected sources in the neighborhood of
our surveys, and using corrections described in SSR, we derived the average
density distribution for the ADBS shown in Fig.~3a. We found that outside of
27$^\circ$ from Virgo, the average density at each redshift was relatively
uniform, while within this radius there was a strong peak in density near the
Virgo redshift. The figure also includes the density distributions for the two
previous Arecibo surveys (SSR), which both show significant peaks at the
Pisces-Perseus redshift.

\begin{figure} 
\plottwo{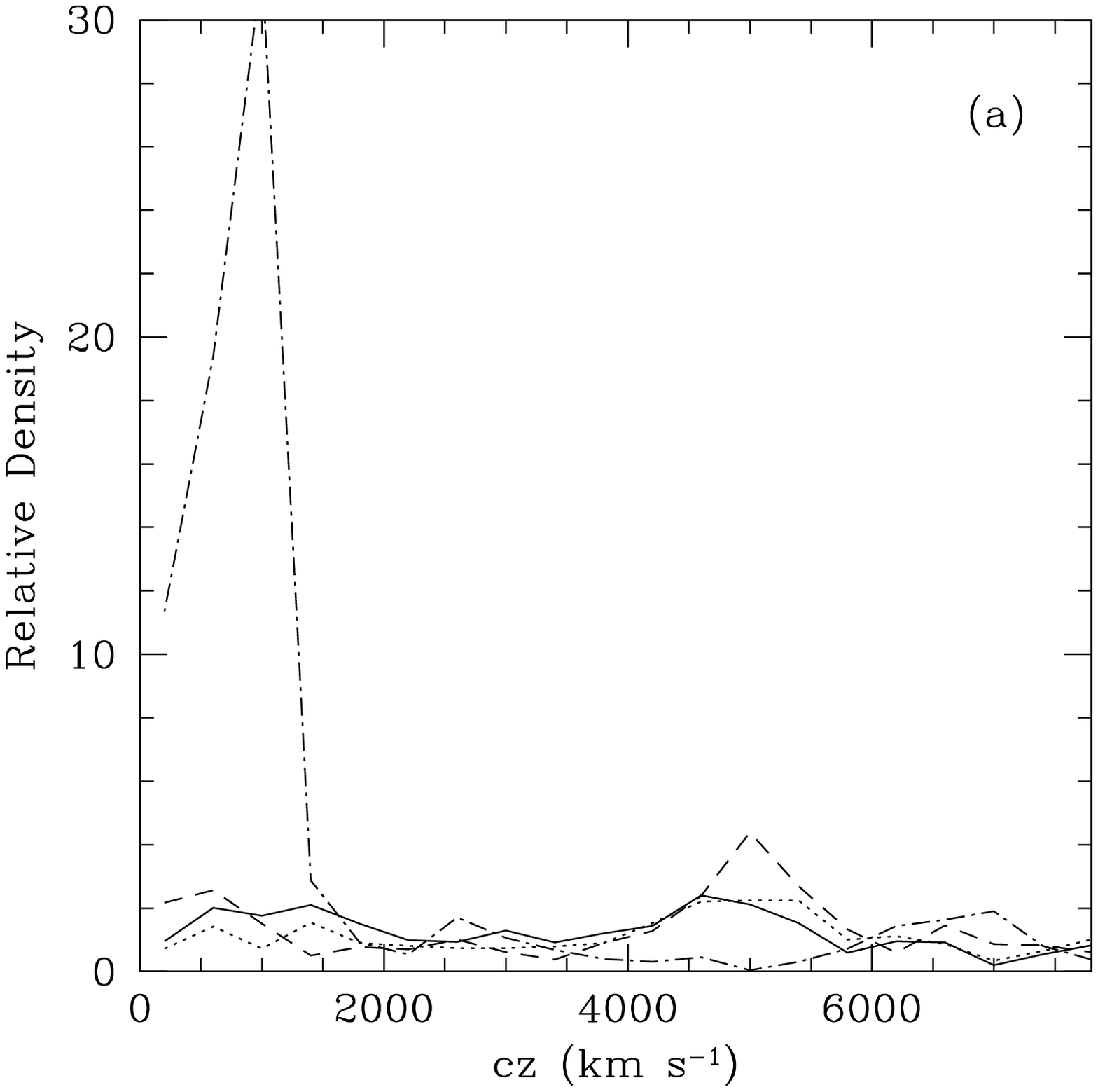}{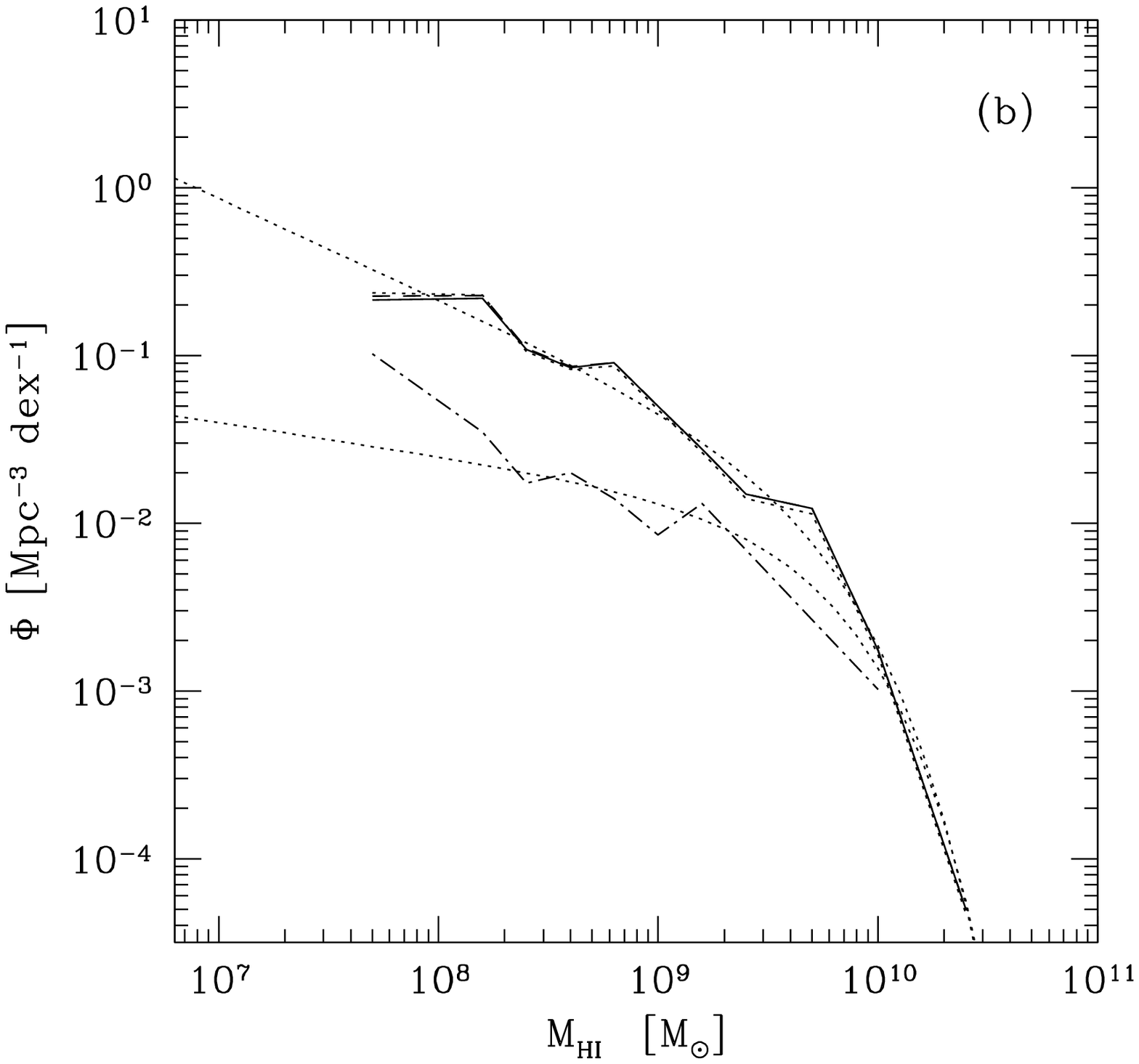} 
\caption{(a) The density structure in the vicinity of the Arecibo blind surveys
based on optical catalogs of galaxies. The solid and dot-dash lines show the
results for the field and Virgo ADBS samples. The dotted and dashed lines are
for the Zwaan et al.~and Spitzak \& Schneider samples respectively. (b) The
field and Virgo HI mass functions. The Virgo function is shown as a dot-dash
line, while three separate curves are shown for the field mass function based
on differing assumptions: correcting for LSS or not, and showing the effect of
applying a different low-velocity cutoff (see text). Model Schechter functions
are shown as in Fig.~2a.}
\end{figure}

The resulting mass function inside and outside the Virgo cluster is shown in
Fig.~3b. Inside the Virgo region, the uncorrected mass function has a higher
density than for the field sample, but the LSS corrections in 
this region drive the corrected density of sources below the field mass
function. The Virgo mass function appears to be consistent with a
shallow slope of the Schechter function.

By contrast, because of the relatively uniform density in the field sample
region, the LSS correction has almost no effect on the final mass function. Two
nearly indistinguishable curves show the field mass function with the LSS
density correction turned on and off. We also show the effect of changing the
minimum redshift cutoff from 100 to 300 km s$^{-1}$; the result is, again,
almost indistinguishable.

Another concern is the effect of random velocity errors on the mass function.
Because of the effects of local dynamical motions and large scale flows, our
use of the Hubble law to derive galaxy distances generates an uncertainty in
the distances and masses. Because of the larger volume at larger distances, one
might fear that this effect would scatter more sources to smaller distances
than vice versa, and therefore bias the lowest-mass bins. However, if the
Schechter slope is steeper than $\alpha=-1.0$ (which every analysis
agrees with), the effect is reversed---the larger number of low mass
sources overcomes the volume difference and the low-mass bins should
be depressed. We have tested this with simulations, and confirm this
conclusion.

We have also tested the effect of using alternative distances based on the
POTENT model of local large scale flows (Bertschinger et al.~1990). This model
shifted points slightly, but well within the error bars shown in the derived
mass functions.

\section{Discussion}

The difference seen in and out of the Virgo Cluster region suggests a strong
environmental dependence of the HI mass function. It is perhaps not surprising
that low-mass HI sources are rare in a cluster where a deep gravitational
potential is needed to retain a galaxy's gas against the effects of
ram-pressure stripping. This might explain why the mass function appears much
more similar in shape to the luminosity function of optically-selected
galaxies.

The shallow mass function seen in moderate-density groups and clusters may be
explained by related effects. For example, even though Ursa Major has a long
crossing time and no evidence of a hot intracluster medium, the fact that the
mean density is an order of magnitude higher than the field implies that the
region has been locally converging with respect to the Hubble flow. This in
turn suggests that the rate of interactions and mergers between low-mass HI
sources and nearby galaxies will be greatly enhanced, again suppressing the
low-mass end of the mass function.

A density dependence of the HI mass function could be important for the earlier
Arecibo surveys. They covered a small area of the sky, and are therefore
subject to localized differences in the HI mass function. In particular, both
derived a substantial fraction of their high-mass counts from the
Pisces-Perseus supercluster region, while the low-mass sources were detected
nearby in regions where the density of sources is more similar to the norm for
field galaxies. This contrast in LSS density for different mass sources might
explain the differences from the ADBS survey, which covered a much larger area
of the sky more shallowly. As a result the ADBS survey volume (excluding the
Virgo Cluster) is more representative of the field.

The effects of large scale structure do not explain all of the differences
between surveys, however. We comment again on the importance of producing
testable measures of the completeness of HI surveys. There are too many subtle
effects in the data-processing stream to simply declare a limit based on what
is expected or detected. As we have shown, some low $S/N$ sources are likely to
creep into a sample and may suggest a better sensitivity than is warranted.
There are also clearly peculiar effects on the completeness as a result of 
source linewidth that may differ depending on the precise analysis procedure.

An additional effect that we have not explored, but which may be especially
important for higher-resolution synthesis mapping, is the effect of source
size. For example, many of the low-mass sources detected in the ADBS appear to
have much larger HI diameters than their optical sizes; some appear extended
even relative to the Arecibo beam. Spreading the emission over such a large
area will suppress such sources' detectability in synthesis mapping, but may
enhance the chance of their detection in sparsely sampled surveys like the
ADBS.

The competing effect of confusion mainly influences single-dish measurements of
the mass function. Since the two-point correlation function indicates that
galaxies tend to cluster around other galaxies, the challenge for a large-beam
21 cm instrument is to separate sources and determine which ``should have
been'' detected based on the survey sensitivity.

It will do little good to conduct larger surveys if the potential biases caused
by selection effects are not well understood. The good news is that with modern
computational resources, it is relatively simple to model false sources and to
insert them into raw survey data. Effects of source size and confusion can be
established directly, and then a more useful comparison of results can be
carried out.

\end{document}